\documentclass[aps,preprint,prd,epsfig]{revtex4}
\usepackage{amssymb}
\usepackage{amsmath}
\usepackage{graphicx}
\usepackage{epsfig}
\usepackage{ccaption}
\usepackage{color}
\usepackage{fancyhdr}
\newcommand{\be}{\begin{equation}}
\newcommand{\ee}{\end{equation}}
\newcommand{\bea}{\begin{eqnarray}}
\newcommand{\eea}{\end{eqnarray}}
\newcommand{\bes}{\begin{subequations}}
\newcommand{\ees}{\end{subequations}}

\newcommand{\bc}{\begin{center}}
\newcommand{\ec}{\end{center}}
\begin{document}
\title{  Inflation scenario driven by a low energy physics  inflaton}
\author{J. G. Ferreira Jr, C. A. de S. Pires, J. G. Rodrigues , P. S. Rodrigues da Silva}
\affiliation{{ Departamento de F\'{\i}sica, Universidade Federal da Para\'\i ba, Caixa Postal 5008, 58051-970,
Jo\~ao Pessoa, PB, Brazil}}

\date{\today}

\begin{abstract}
It is a longstanding desire of cosmologists, and particle physicists as well, to connect inflation to low energy physics, culminating, for instance, in what is known as Higgs inflation. The condition for the standard Higgs boson playing the role of the inflaton, and driving sucessfully inflation,  is that  it couples  nonminimally  with gravity.  Nevertheless,  cosmological constraints impose that  the nonminimal coupling be large. This causes the loss of perturbative unitarity in a scale of energy far below the Planck one. Our aim in this work is to point out that inflaton  potential containing  a particular  type of trilinear coupling involving the inflaton  may circumvent this problem by realizing Higgs inflation with tiny nonminimal coupling of the inflaton with gravity. We then develop the idea within a toy model and implement it in the inverse  type-II seesaw mechanism for  small neutrinos masses. 
 \end{abstract}
\maketitle
\section{Introduction}
\label{secint}
With the advent of precision measurements concerning the power spectrum of the cosmic microwave background radiation (CMB)~\cite{Bennett:2012zja}\cite{Ade:2015tva}\cite{Ade:2015lrj}, the inflationary paradigm\cite{Guth:1980zm}\cite{Linde:1981mu}\cite{Albrecht:1982wi} has conquered the status of a feasible physical theory that can be put on trial. More specifically, some classes of models can already be discarded when contrasted with cosmological inflation parameters as bounded by the Planck satellite~\cite{Ade:2015lrj}. This is a remarkable achievement once one considers that inflation was hardly taken seriously a few decades ago. After much development, theoretical and experimental, in the physics of the inflationary universe, there are mainly two points that keep challenging people who struggle to make it a robust and, possibly, testable theory at the energy scale probed at the LHC, for example. First, it is not generally easy to reconcile the tiny amplitude of the power spectrum of density perturbations  with field theoretical inflationary models without fine-tuning\cite{Linde:2005ht} and, secondly, it is not clear how  to entertain a natural scenario where inflation might be connected to the low energy physics\cite{Ellis:2016qru}.

Concerning the first point,  models capable of, somehow, connecting  inflation to the low energy particle physics  inevitably must involve a quartic self-coupling in the inflaton potential, $\lambda \phi^4$,  which is severely constrained by the  smallness of the amplitude of curvature perturbation. The difficulty here is  finding an inflaton model belonging to the low energy scale  that can fulfill the   phenomenological inflationary constraints. For example, the  main candidate to play the role of the inflaton at the low energy scale is  the  standard Higgs itself.  However, this case is  already discarded since a  mass of $125$~GeV for the Higgs boson requires $\lambda=0.6$.  This value for the quartic coupling does not fit  the cosmological observables. For the same reasons, even the  straightforward  extension of the SM potential with  new scalar fields  do not guarantee successful inflation. 

A way of circumventing these drawbacks  is to allow the large nonminimal coupling between the Higgs boson and gravity\cite{Makino:1991sg}\cite{Bezrukov:2007ep}. This approach, albeit being an appealing inflation mechanism,  has been fairly criticized because it suffers from loss of perturbative unitarity at energies far below the Planck scale~\cite{Barbon:2009ya}, although there is some dispute about this being a real obstacle~\cite{Escriva:2016cwl}. In view of this, in what concerns the link of inflation to low energy physics, we seem to be at square one again.

In this work we point out that potentials that allow  a particular type of trilinear coupling may conciliate inflation with low energy scale for the case of tiny nonminimal coupling of the inflaton with gravity. Such term is the main ingredient of type-II seesaw mechanism. Thus, as a concrete example, we consider the standard model added by a scalar triplet characterizing the type-II seesaw mechanism for small neutrino mass. We show that when lepton number is assumed to be explicitly violated at very low energy scale (inverse type-II seesaw mechanism), neutrinos develop mass at eV scale and the neutral component of the scalar triplet, which develops mass around TeV scale, may play the role of the inflaton and drives inflation even for the case of a tiny nonminimal coupling of the inflaton with gravity.

The paper is divided in the following way: In  Sec.~\ref{sectoy} we present the idea by making use of a toy model. Next, in Sec.~\ref{secmodel}, we develop  a real model based on a triplet extension of the standard model.  We present our conclusions in Sec.~\ref{conc}.

\section{Inflation models belonging to the  electroweak scale: challenges and solutions}
\label{sectoy}
Suppose inflation is driven by a scalar field, $h$, whose potential is,
\be 
V(h)= \mu_h^2 h^2  +\lambda_h h^4.
\label{higgspot}
\ee
In order for this potential to fit the phenomenological inflationary constraints in the form of small slow-roll parameters ($\epsilon , \eta \ll 1$) and cosmological perturbations ($\Delta^2_R=2.215\times 10^{-9}$), we need to choose $\lambda_h \approx 10^{-12}$. This is in direct contradiction to the observed Higgs mass at the LHC, being the reason why the standard Higgs boson cannot drive inflation once its self-coupling is already fixed at $\approx 0.6$. Another problem with Higgs inflation concerns the vacuum stability of its  potential that is stable up to energies around $10^{11}$~GeV, this is particularly troublesome for inflation that involves energies above the Planck scale. 

Suppose now that there are two scalar fields $h$ and $\sigma$ composing the following potential:
\be 
V(h,\sigma)= \mu_h^2 h^2 + \mu_\sigma^2 \sigma^2 +\lambda_h h^4 + \lambda_\sigma \sigma^4 +\lambda_{\sigma h}h^2 \sigma^2.
\label{toypot}
\ee
This potential may represent an extension of the Higgs sector of the SM by a singlet scalar $\sigma$.  If one assumes that inflation occurs in the $\sigma$ direction then, the term that matters during inflation is $\lambda_\sigma \sigma^4$, and the inflaton mass is given by $m_\sigma \approx \sqrt{\lambda_\sigma}v_\sigma$ where $v_\sigma$ is the vacuum expectation value (vev) of $\sigma$.  Nevertheless, as we have already mentioned, the constraints from CMB imply  $\lambda_\sigma \approx 10^{-12}$, and if we wish to connect inflation to low energy physics, imposing $v_\sigma$ at electroweak scale, we see that the inflaton acquires a very tiny mass, jeopardizing the reheating phase.  The vacuum stability problem may be softened by the mixing coupling $\lambda_{\sigma h}$ though. Then, in summary, the smallness of inflaton quartic coupling, tuned to produce the observed amplitude of curvature perturbation, is the constraint that seems to impede inflation to be implemented at the electroweak scale.

On the other hand, by allowing a non-minimal coupling between the Higgs boson and gravity as follows,
\be
{\cal L} \supset  {\cal L}_{SM} -\frac{M_P^2}{2}R -\xi h^*hR,
\label{nonminimal}
\ee
where $M_P$ is the reduced Planck mass and $R$ is the Ricci scalar, one may have the SM Higgs boson driving successful inflation~\cite{Bezrukov:2007ep}. However, the  price to pay for this is that   $\xi$ must take large values, more specifically around $10^4$. The unpleasant consequence of this result is that perturbative unitarity is lost at an energy scale dictated by $\frac{M_P}{\xi}$, lower than the Planck scale by 4 orders of magnitude.  Such a proposal has provoked an intense discussion as to whether such theories make sense when quantum corrections are taken into account, since $\xi > 1$~\cite{Barbon:2009ya}\cite{Escriva:2016cwl}\cite{Hertzberg:2010dc}\cite{Lerner:2009na}. Without taking part in this issue, this approach encouraged the search for models that connect inflation to the electroweak scale physics. 

In this sense, we discuss what an extension of the standard Higgs potential might contain in order to posses an inflaton that belongs to low energy scale and drives successfully inflation. 

Before going to a realistic scenario, let us consider the potential in Eq.(\ref{toypot}) augmented by the trilinear term $\mu h^2 \sigma$.  During inflation the $\lambda_\sigma \sigma^4$ term dominates.  The  amplitude of curvature perturbation still requires a very small $\lambda_\sigma$, but now its influence on the inflaton mass is very small once  the trilinear  term contributes to the mass of $\sigma$ by  $m^2_\sigma \sim \lambda_\sigma v_\sigma^2 +\frac{\mu}{v_\sigma}v_h^2 \rightarrow m_\sigma \sim \sqrt{\frac{\mu}{v_\sigma}}v_h$.  For $\mu$ around $v_\sigma$, the inflaton mass is around $v_h$, the vev of $h$,  regardless  of  $\lambda_\sigma  \ll1$.    In the next section we discuss a realistic implementation of this idea in a rather phenomenological model where such a trilinear term is necessary in order  to generate the seesaw mechanism for small neutrino masses.

\section{Inflation in the TeV type-II seesaw mechanism}
\label{secmodel} 
Let us consider the SM augmented by the scalar triplet $\Delta$,
\begin{equation}
\Delta\equiv \left(\begin{array}{cc}
\frac{\delta^{+}}{\sqrt{2}} & \delta^{++} \\ 
\delta^{0} & \frac{-\delta^{+}}{\sqrt{2}}
\end{array} \right)\,\sim (1, 3,2), 
\end{equation}
which has  hypercharge, $Y=2$ and lepton number, $L=-2$.  The scalar potential can be written as
\begin{eqnarray}
V(H,\Delta) &=& -m_{H}^{2}H^{\dagger}H + \frac{\lambda}{4}(H^{\dagger}H)^{2} + M^{2}_{\Delta}Tr[(\Delta^{\dagger}\Delta)]+[\mu(H^T i\sigma^{2}\Delta^{\dagger}H)+H.c]\nonumber \\
&&+\lambda_{1}(H^{\dagger}H)Tr[(\Delta^{\dagger}\Delta)] +\lambda_{2}(Tr[(\Delta^{\dagger}\Delta)])^{2} +\lambda_{3}Tr[(\Delta^{\dagger}\Delta)^{2}] \nonumber \\
&&+ \lambda_{4}H^{\dagger}\Delta^{\dagger}\Delta H + \lambda_{5}H^{\dagger}\Delta\Delta^{\dagger}H\,, 
\label{potential}
\end{eqnarray}
where $H = (H^{+}\,\,\,\, 
H^0)^T\sim(1,2,1)$ is the standard Higgs doublet.
The trilinear term explicitly violates lepton number with the parameter $\mu$ being the energy scale associated to such violation. 
The neutral components of $H$ and  $\Delta$ are shifted in the conventional way $
 H^{0} , \Delta^{0} \rightarrow \frac{1}{\sqrt{2}}\left(  v_h + h +i I_h , v_\delta
+ \delta +iI_\delta\right)$. 

Consider the lepton number violating scale $\mu$. Although it can take any value,  it seems that people tend to assume high values. However, there is no apparent reason to not stick to low values too. Here we follow the  t' Hooft naturalness argument which says that a small $\mu$ is more natural than a high one because symmetry is enhanced when $\mu \rightarrow 0$\cite{tHooft:1979rat}. In this case we refer to such a mechanism as the inverse type II seesaw one\cite{Freitas:2014fda}.

In what concerns neutrino physics, the importance of the trilinear term in the potential above reveals in the relation $ v_\delta  \simeq \frac{1}{\sqrt{2}} \frac{v^2_h}{M^2_\Delta}\mu $\  provided by  the minimum conditions of the potential in Eq.~(\ref{potential})~\cite{Cheng:1980qt}\cite{Arhrib:2011uy}\cite{Freitas:2014fda}. Moreover, with this scalar triplet we have the  Yukawa interactions  $Y_{ij}\bar{f}^{c}_{i} i \sigma_{2}\Delta f_{j}$ where  $f=(\nu \,\,,\,\, e)_L^T$. Thus, when $\delta^0$ develops a vev different from zero neutrinos gain Majorana mass $m_\nu =\frac{1}{\sqrt{2}}Yv_\delta=\frac{1}{2}Y \frac{v^2_h}{M^2_\Delta}\mu $. A small $v_\delta$ guarantees small neutrino masses. What is more interesting here is that because we have $v_\delta \sim$~eV, and consequently have neutrinos at eV scale,  we need as input   $M_\Delta \sim $~TeV, $\mu \sim$~keV and  $v_h \sim 10^2$~GeV. In other words, this model provides neutrino mass at eV scale with new physics at TeV scale~\cite{Arhrib:2011uy}\cite{Freitas:2014fda}.  Thus, if somehow  the potential above performs inflation, then we may say that it connects inflation to low energy physics.

Because of the trilinear term the $\delta^0$ mass provided by the potential above is given by $m^{2}_{\delta}\sim (\frac{\lambda}{4}+ \frac{1}{\sqrt{2}}\frac{\mu}{v_{\delta}})v^{2}_{h}$ and the mixing between $\delta^0$ and $h$ is of the order of $\sqrt{\frac{v_\delta}{v_h}} \sim 10^{-6}$, revealing that $\delta^0$  decouples from $h$. 


We now address inflation in the context of this model. We remark that in this model too, the Higgs boson, $h$, cannot play the role of the inflaton, given that its mass is predominantly dependent on $\lambda$, which would be constrained to extremely small values when considering the amplitude of primordial density perturbations. This is in disagreement with the large value $\lambda$ has to assume so as to recover the observed Higgs mass. However, the $\delta^0$ is at hand and, as we are going to show, it is perfectly suitable to incarnate the inflaton field, and that is the direction we choose for inflation to happen. For previous works with scalar triplet inflation,  see \cite{Chen:2010uc}\cite{Arina:2012fb}. In the first, the scalar triplet belongs to GUT scale. The second considers heavy triplet and inflation scenario involving large nonminimal coupling of the inflaton with gravity. Our case is different from both of these even in the form of the potential during inflation and in the phenomenology of the scalar triplet at low energy scale.

During inflation, the relevant piece of the scalar potential is $V(\delta)\approx \frac{\lambda_2 +\lambda_3}{4} \delta^4$. One can be optimistic with this model since we promptly see that $\delta^0$ mass, $m^{2}_{\delta}\sim (\frac{\lambda}{4}+ \frac{1}{\sqrt{2}}\frac{\mu}{v_{\delta}})v^{2}_{h}$, does not depend on the tiny couplings, $\lambda_2$ and $\lambda_3$, constrained to fit the amplitude of curvature perturbation.

For the set of values we already displayed above, $\mu\approx 0.1$~keV, $v_\delta=1$~eV and $v_h= 10^2$~GeV, we obtain $m_\delta \approx$~TeV, which is very appropriate for generating an efficient reheating phase. Notwithstanding, at this stage, this model is unable to yield neither the observed spectral index $n_s$ nor the tensor to scalar ratio $r$, as extracted from Planck 2015 results~\cite{Ade:2015lrj}.
A way out is to either consider radiative corrections to the potential\cite{NeferSenoguz:2008nn} or nonminimal coupling of $\delta$ with gravity . In what follows we develop the second option.
 
Although  both,  $h$ and $\delta$,  may couple nonminimally to gravity according to  $\xi_h h^{\dagger} h R + \xi_\delta \delta^{\dagger} \delta R$, for simplicity we suppose that inflation happens along the $\delta$ direction only. In practical terms this means $\xi_h =0$ and that large field values (Planckian) are assumed by $\delta$ only.  In this case our Lagrangian of interest in the Jordan frame  contains the terms,
 \be
 {\cal L} \supset \frac{1}{2} (\partial_\mu \delta)^{\dagger}(\partial^\mu \delta)-\frac{M_P^2R}{2}-\frac{1}{2}\xi_\delta \delta^2 R -V(\delta),
 \label{Ljordan}
 \ee
where $V(\delta)=\frac{\lambda_2 +\lambda_3}{4} \delta^4$.  The Jordan frame is the physical one, but we can go to the Einstein frame, by a Weyl transformation, where the dynamics of inflation is simpler to investigate.
Such a transformation involves the relations\cite{Faraoni:1998qx}
\bea
\tilde g_{\mu \nu}&&=\Omega^2g_{\mu \nu}\,\,\,\,\,\,\mbox{where}\nonumber \,\,\,\,\,  \Omega^2= 1+\frac{\xi_\delta \delta^2}{M^2_P}\nonumber \\
&& \frac{d\chi_\delta}{d\delta}=\sqrt{\frac{\Omega^2 +6\xi^2_\delta \delta^2/M_P^2}{\Omega^4}}.
\label{conformaltransf}
\eea
In the Einstein frame the Lagrangian should contain the following terms,
\be
 {\cal L} \supset -\frac{M^2_{P} \tilde R}{2}+\frac{1}{2} (\partial_\mu \chi_\delta)^{\dagger}(\partial^\mu \chi_\delta)-U(\chi_\delta)\,,
 \label{LEinstein}
 \ee
where $U(\chi_\delta)=\frac{1}{\Omega^4}V(\delta[\chi_\delta])$. In order to obtain the potential $U$ in terms of $\chi_\delta$ we write $\delta$ as a functional of $\chi_\delta$, by integrating Eq.~(\ref{conformaltransf}) and substituting the result in $V({\delta[\chi_\delta]})$. For the case where $\xi_\delta >0$, we have\cite{Accioly:1993kc}\cite{GarciaBellido:2008ab}
\bea
\chi_\delta&=&\sqrt{\frac{1+6\xi_\delta}{\xi_\delta/M^2_P}}\ln\left[\sqrt{1+\xi_\delta(1+6\xi_\delta)\frac{\delta^2}{M^2_P}} +\sqrt{\xi_\delta(1+6\xi_\delta)\frac{\delta^2}{M^2_P}}\right] \nonumber \\
&&-\sqrt{\frac{3}{2}M^2_P}\ln\left[\frac{\left(  \sqrt{1+\xi_\delta(1+6\xi_\delta)\frac{\delta^2}{M^2_P}} +\sqrt{6 \xi^2_\delta \frac{\delta^2}{M^2_P}} \right)^2}{1+\xi_\delta\frac{\delta^2}{M^2_P}}   \right].
\label{relatingfields}
\eea
We restrict ourselves to the situation where  $\xi_\delta \ll 1$, thus simplifying the above expression\cite{Ballesteros:2016euj}.  In this case the potential as a function of $\chi_\delta$ is given by
\begin{equation} U\left(\chi_\delta\right)\approx \left( \frac{\lambda_2 + \lambda_3}{4(1+3\xi_\delta)^4} \right)\left( \frac{\chi_{\delta}^{4}}{\left( 1+ 
 \frac{\xi_\delta \chi_{\delta}^{2}}{M^2_P (1 + 3\xi_\delta)^2}\right)^2} \right).
 \label{eisteinframe}
\end{equation}
This potential is not exponentially flat as in the case of large $\xi_\delta$ considered in \cite{Arina:2012fb}.  Surprisingly, it is flat enough to get in accordance with the phenomenological inflationary constraints in the form of  $\Delta_R^2$,  $n_s$ and $r$ as imposed by recent data from Planck2015.  Perceive that for small values of $\chi_\delta$, the Jordan field coincides with the Einstein one and the potential recovers the canonical  form of Eq.~(\ref{potential}). 

Let us now treat the issue of inflation that takes place  when the slow roll approximation is satisfied, ($\epsilon \ll 1$, 
$\eta \ll 1$), where~\cite{Liddle:1994dx}
\begin{eqnarray}
 &\epsilon & = \frac{M^2_{P}}{2}\left(\frac{ U^{\prime}}{ U}\right)^2, \quad \quad
 \eta  = M^2_{P}\left( \frac{U^{\prime \prime}}{U}\right),
\end{eqnarray}
with $M_P=2.4 \times 10^{18}$~GeV. 

The spectral index $n_S$ and
the scalar to tensor  ratio $r$  are defined as\cite{Lyth:2009zz}
\begin{eqnarray}
 &n_S&=1-6\epsilon+2\eta, \quad \quad \quad r=16\epsilon, \nonumber \\
\end{eqnarray}

For a wave number $k=0.05$~Mpc$^{-1}$, the Planck results indicate $n_S=0.9644 \pm 0.0049$ and $r<0.149$

The number of e-folds is given by
\begin{equation}
N = \frac{-1}{M^2_P}\int_{(\chi_\delta)_i}^{(\chi_\delta)_f}\frac{U}{U^{\prime}}d\chi_\delta,\label{efold}
\end{equation}
where $(\chi_\delta)_{f}$ marks the end of inflation and is defined by $(\epsilon,\eta)=1$. Its value is $(\chi_\delta)_{f}=8.30\times 10^{18}$~GeV.
To find $(\chi_\delta)_i$ we set  $N=60$ and solve Eq.~(\ref{efold})  for $(\chi_\delta)_i$.  Its value is $(\chi_\delta)_{i}=5.26\times 10^{19}$~GeV.

Another important parameter is the amplitude of curvature perturbation $\Delta^2_R = \frac{U}{24\pi^2 M_P^4\epsilon}$. Planck measurement of this parameter gives $\Delta^2_R=2.215\times 10^{-9}$ for a wave 
number $k=0.05$~Mpc$^{-1}$. This imposes  tiny values for the parameters $\lambda_1$, $ \lambda_2$, $\lambda_3$, $\lambda_4$ and $\lambda_5$. 

In FIG.~1 we present our results by plotting  $r$ versus $n_s$ generated by  the potential $U(\chi_\delta)$. We varied $\xi_\delta$  from $10^{-4}$ up to $10^{-2}$ for $N=50(\mbox{green line})\,,\,55(\mbox{blue line})\,,\,60(\mbox{red line})$. Perceive that for $N=60$ and  $\xi_\delta =10^{-2}$ the model predicts $r \sim 0.05$.  In general, the larger $\xi_\delta$  is, the smaller  $r$ gets. For previous work of the Higgs-inflation scenario considering  $\xi \leq 1$, see \cite{Ballesteros:2016euj}. It is interesting to notice that even for $\xi_\delta \ll1$ the nonminimal coupling still plays an important role  in providing a better accuracy of the spectral index and the tensor to scalar ratio than inflation driven by  a potential with quartic coupling and minimal coupling to gravity. This result is encouraging in the sense that the low energy inflation model may arise from gauge theory where the scalar potential has the general  form of  Eq.~(\ref{potential}). We know from recent results that inflaton  potential composed exclusively by  quartic coupling is practically excluded\cite{Ade:2015lrj}. However when we consider this potential but now with a nonminimal coupling between the inflaton and gravity, we see that acceptable inflation is achieved even for tiny non-minimal coupling. This remarkable outcome allows us to connect inflation to low energy physics within models where the inflaton mass does not depend dominantly on the quartic coupling anymore.  
For a supersymmetric model that can lead to successful inflation along the
same line, see \cite{Kasuya:2003iv}

\begin{figure}[h]
\centering{\includegraphics[scale=0.4]{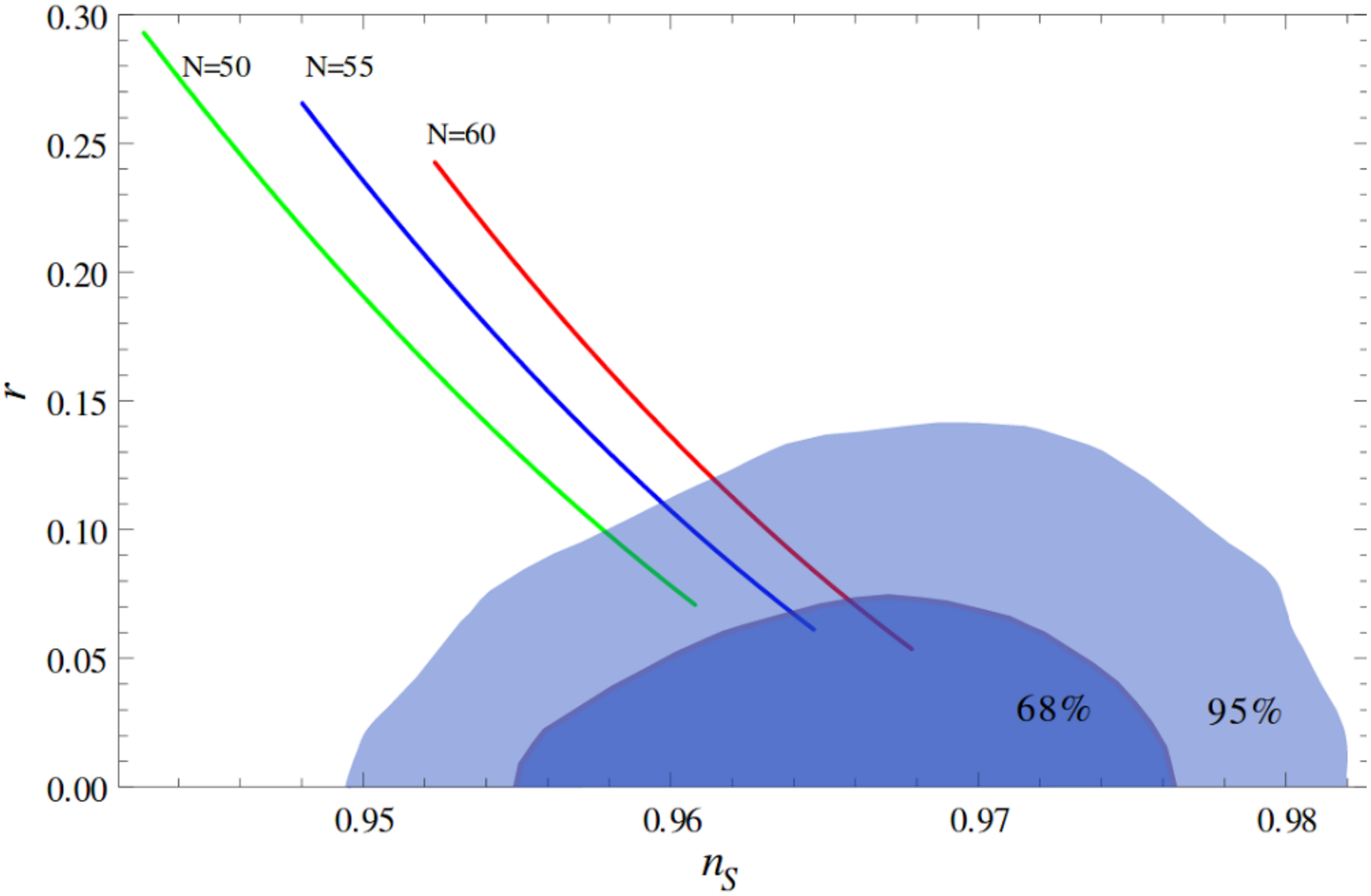}}
\caption{ Triplet inflation predictions  for $n_S$ and $r$  with $\xi_\delta$ varying from $10^{-4} $ up to $10^{-2}$. The region in blue is the Planck15 data  at the 68\% and 95\% CL }
\label{nsversusr}
\end{figure}

After the inflationary phase, the inflaton  oscillates around its vev giving rise to the reheating phase\cite{Abbott:1982hn}\cite{Allahverdi:2010xz}\cite{Linde:2005ht}. In canonical inflation, reheating is achieved by the decay of the inflaton into pairs of standard particles. In the presence of nonminimal couplings, be it tiny or large, things are significantly more complicated and numerical study in lattice is made necessary. This is so because during reheating  perturbative and nonperturbative effects\cite{Kofman:1997yn} are mixed. Here, following Ref. \cite{GarciaBellido:2008ab},  we just argue why  perturbative decay of the inflaton in pair of standard particles  is not instantaneous as in the canonical inflation case. The two conditions for perturbative decay of the inflaton are:  enough phase space, which translates in $M_{\chi_\delta} > 2m_f$ and that the decay rate, $\Gamma_{\delta \rightarrow 2f}$, be greater than the rate of expansion $H=\sqrt{\frac{\rho_{\chi_\delta}}{3M_P^2}}$.  For we have an idea of the complexity of the  situation, for the case of the standard Higgs as the inflaton, the Higgs condensate must oscillate $10^{12}$ times before the decay rate into electrons overtakes the Hubble rates\cite{GarciaBellido:2008ab}. In our case, the triplet $\Delta$ generate mass exclusively to the standard neutrinos. Thus $m_\nu= \frac{Y \chi_\delta}{\sqrt{2}(1+3\xi)}$ while $M_\delta= \frac{M_\Delta}{(1+3\xi)}$. For $(\chi_\delta)_{end}=8.3 \times 10^{18}$ GeV we get $m_\nu \gg M_\delta$. Thus, our case is not different from the other Higgs inflation scenarios and numerical studies in lattice will also be necessary.

\section{conclusions}
\label{conc}
To connect inflation with low energy physics is a longstanding desire of cosmologists and particle physicists as well. The quest for a mechanism capable of  linking inflation with phenomenological particle physics models has received extensive attention lately. That is so since the proposal  that large nonminimal coupling of scalars with gravity may allow  that the standard Higgs boson plays the role of the  inflaton.  However, there is some discomfort with such an idea because perturbative unitarity is violated at energy  scale $\Lambda$ below the Planck one. A solution to this problem would involve new physics at $\Lambda$ scale\cite{Giudice:2010ka}\cite{Hertzberg:2011rc}. In what concerns vacuum stability and Higgs inflation, we drive the readers to the works \cite{Salvio:2013rja}\cite{Bezrukov:2014ipa}

In practice the difficulties are to conciliate  an inflaton belonging to low energy physics with a very small inflaton quartic self-coupling, as required by the amplitude of curvature perturbation. This is so because in the majority of the models the inflaton mass depends dominantly on the  quartic self-coupling.  A way out of this problem is to look for inflation models whose potentials contain a trilinear term composed by the inflaton and the Higgs boson. This avoids  the inflaton mass getting dominantly dependent on the quartic coupling. We presented the idea, first, in a toy model and next  developed it in a realistic scenario where inflation is implemented within the inverse type-II  seesaw mechanism. In this model the inflaton gains mass around the TeV scale. Thus, by a simple extension of the standard model through a  Higgs triplet belonging to TeV scale  we may  explain the smallness of the neutrino masses and realize inflation with a inflaton connected to low energy physics  in a model that may be probed at the LHC\cite{Freitas:2014fda}. 

In what concerns vacuum stability and Higgs inflation, we drive the readers to the works \cite{Salvio:2013rja}\cite{Bezrukov:2014ipa}. The triplet extension of the standard model helps in solving the problem of the vacuum stability of the standard Higgs\cite{Haba:2016zbu}. However, we are not sure if this result is valid here because the portal couplings  $\lambda_1$, $\lambda_4$ and $\lambda_5$ contribute radiatively to the inflaton potential according to the Coleman-Weinberg approach. On adjusting these couplings to accommodate the amplitude of curvature perturbation,  it may be that they have to develop  values  that are not in accordance with  the solution to the  problem of the vacuum stability  of the standard Higgs boson.  However, this requires a deep investigation of radiative corrections to the  inflaton potential for the case  of small $\xi$.   For studies of radiative corrections for the case of large $\xi$in Higgs inflation and general, see Refs. \cite{Salopek:1988qh}\cite{Barvinsky:1998rn}\cite{Barvinsky:2008ia}

\acknowledgments
This work was supported by Conselho Nacional de Pesquisa e
Desenvolvimento Cient\'{i}fico- CNPq (C.A.S.P,  P.S.R.S ) and Coordena\c c\~ao de Aperfei\c coamento de Pessoal de N\'{i}vel Superior - CAPES (J.G.R). 

\bibliography{inflationtriplet.bib}
\end{document}